\documentstyle[preprint,tighten,aps]{revtex}

\draft 

\begin{document}

\title{ Edge Tunneling of Vortices in Superconducting Thin Films }

\author{ Roberto Iengo }

\address{ International School for Advanced Studies, Via
Beirut 4, 34014 Trieste (Italy) \\
and INFN, Sezione di Trieste, 34100 Trieste (Italy) }

\author{ Giancarlo Jug }

\address{ INFM and Istituto di Scienze Matematiche, Fisiche e Chimiche \\
Universit\`a di Milano a Como, Via Lucini 3, 22100 Como (Italy) \\
and INFN, Sezione di Pavia, 27100 Pavia (Italy) }

\date{ 12 June 1996 } 

\maketitle

\abstract{
We investigate the phenomenon of the decay of a supercurrent due to the
zero-temperature quantum tunneling of vortices from the edge in a thin 
superconducting film in the absence of an external magnetic field. An 
explicit formula is derived for the tunneling rate of vortices, which are 
subject to the Magnus force induced by the supercurrent, through 
the Coulomb-like potential barrier binding them to the film's edge. Our
approach ensues from the non-relativistic version of a Schwinger-type
calculation for the decay of the 2D vacuum previously employed for 
describing vortex-antivortex pair-nucleation in the bulk of the sample. 
In the dissipation-dominated limit, our explicit edge-tunneling formula 
yields numerical estimates which are compared with those obtained for 
bulk-nucleation to show that both mechanisms are possible for the decay 
of a supercurrent.
}

\section{ Introduction }
\renewcommand{\theequation}{1.\arabic{equation}}
\setcounter{equation}{0}

\noindent

Recently there has been much renewed interest in the physics of vortices
in high-temperature superconductors in the presence of applied magnetic 
fields \cite{reviews}. Phenomena involving vortex dynamics and their
hindrance by means of added pinning centres pose well-defined physics
problems of great relevance for the technological applications of the new
high-$T_c$ materials. 

Some of the lesser understood issues are related to the problem of the 
residual resistence due to the thermal or quantum motion of the vortices 
in the presence of an externally applied supercurrent, but in the absence
of the applied magnetic field. Even in the absence of an external field, 
vortices can be dragged into the bulk under the action of the Lorentz-like
Magnus force which they experience sitting at the edge of the sample. 
This phenomenon, leading to an electrical resistence in the presence of 
dissipation, has been investigated recently by relatively few authors 
\cite{aoth94,stephen} considering its importance from both the conceptual 
and the practical points of view. In addition, in a recent article 
\cite{ieju95}
we have pointed out how such a residual resistence may arise, in the 
absence of an applied field, also due to the spontaneous homogeneous 
nucleation of vortex-antivortex pairs in the bulk of the sample.
These are created as fluctuations of
the electromagnetic-like Magnus field acting on the quantised vortices and 
antivortices thought of as electron-positron-like pairs. By means of this 
analogy with quantum electrodynamics (QED), we have set up a  powerful 
``relativistic'' quantum field theory (QFT) approach to study vortex 
nucleation in the two-dimensional (2-D) geometry. This study has been 
conducted in the presence of quantum dissipation and for the cases of 
either an harmonic local pinning potential \cite{ieju95} or, more recently
\cite{ieju96}, for a periodic-lattice distribution of pinning centers. Our 
analysis has produced an explicit analytic result, and an estimate of the 
magnitude of the vortex pair-nucleation rate has shown that the effect may 
become experimentally accessible at low temperatures and high enough 
current densities. 

In this article we investigate, by means of our QFT approach, the original
issue of tunneling of quantised vortices from the sample's boundary as an
alternative mechanism for the decay of an applied supercurrent in a 
thin-film geometry. The problem has already been introduced
by Ao and Thouless \cite{aoth94} and by Stephen \cite{stephen}, 
with and without the inclusion of quantum dissipation, and discussions that 
give full weight to the inertial mass of the moving vortex can be found in the
literature \cite{aoth94} alongside treatments \cite{stephen,ivlev} in which 
the inertial mass is taken to be negligible. The physics of the problem has 
been captured mainly through semi-classical evaluations of the tunneling 
rate and discussion has centered on the effects of dissipation and pinning 
in opposing the stabilising influence of the Magnus force on the classical 
orbits of a vortex \cite{aoth94,stephen}. 
Beside the issue of the residual superconductor's
resistence, the quantum tunneling of the magnetic flux lines into the bulk 
of the material has been investigated to understand theoretically 
\cite{ivlev,blatter} the observed \cite{fiorani} flux-creep phenomenon in the 
presence of a field.    

In this work we will analyze further
the problem of vortex tunneling from the point
of view of the instability of the ``vacuum'' represented by a thin 
superconducting film in the presence of an externally-driven supercurrent.
Our aim is in fact that of obtaining an explicit formula (complete of
prefactors) for the tunneling rate. 
From the theoretical point of view, our approach differs from other
semi-classical evaluations of the path integral in that use is made of the 
Schwinger formalism for pair creation in QFT. Since this fully-relativistic 
formulation provides for a description of vortex-antivortex pair nucleation, 
we will be concerned here with the completely non-relativistic limit of this 
approach, formally corresponding to the case of a negligible vortex inertial 
mass and to contributions to the path integral relating to quantum 
particles moving ``forward'' in the time coordinate. This non-relativistic
version of the Schwinger formalism indeed appears as very convenient 
for carrying out explicit and straightforward calculations, especially in the
dissipation-dominated regime. We in fact work out the tunneling rate in the 
case when dissipation dominates over inertia, 
and for the case of a potential barrier made up of the 
Coulomb-like attraction to the edge and the electric-like potential extracting 
the vortex into the bulk. Our central result shows that the tunneling rate 
$\Gamma$ has a strong exponential dependence on the number current 
density $J$, much as in the case of  vortex nucleation in the bulk of the 
sample \cite{ieju95}. The numerical value of the tunneling rate per unit 
length, as obtained from our approach with material parameters typical of 
the high-temperature cuprate superconductors, compares favourably with 
analogous results obtained for the bulk nucleation rate. This qualifies 
the novel vortex nucleation mechanism by us proposed in recent articles 
\cite{ieju95,ieju96} as almost as likely as vortex tunneling from the 
sample's edge for the decay of a supercurrent. In our estimates, edge 
tunneling appears to be more favourable than bulk nucleation for equal 
given extension of edge length and surface area.

It is useful to describe the problem at hand by means of the classical 
equation of motion for a single vortex moving at relatively low velocity in a 
supercurrent having density ${\bf J}$ 

\begin{equation} 
m\ddot{\bf q}=-{\nabla}U({\bf q})+e{\bf E}-e\dot{\bf
q}{\times}{\bf B}-\eta\dot{\bf q} 
\label{classic} 
\end{equation}

\noindent 
Here $m$ is the (negligible) inertial mass of the vortex carrying topological 
charge $e=\pm2\pi$ and treated as a single, point-like particle of 2-D 
coordinate ${\bf q}(t)$. Also, $U({\bf q})$ is the phenomenological potential 
acting on the vortex, in the present case the ``Coulomb-like''  interaction 
binding it to the sample's edge. The supercurrent ${\bf J}$ gives rise to an 
electric-like field ${\bf E}={\times}{\bf J}$ (a notation implying
${\bf E}\cdot{\bf J}=0$) superposed to a magnetic-like field 
${\bf B}=\hat{\bf z}d\rho_s^{(3)}$ for a thin film orthogonal to the vector 
$\hat{\bf z}$ and having thickness $d$ with a superfluid component 
characterised by a 3-D number density $\rho_s^{(3)}$. Finally, $\eta$ is a 
phenomenological friction coefficient taking dissipation into account. A
derivation of this electromagnetic analogy for the Magnus force was given
by us in Ref. \cite{ieju95} (but see also \cite{lefi,niaoth,gaitan}). Further
contributions to ${\bf B}$ arising from other quantum effects are possible
and have been recently debated \cite{ofgb}, however in the 
dissipation-dominated regime the actual explicit dependence of ${\bf B}$
on the material's parameters will be irrelevant for our study. The 
quantum-mechanical counterpart of  Eq. (\ref{classic}) is constructed 
through the Feynman path-integral transposition in which the dissipation is 
treated quantistically through the formulation due to Caldeira and Leggett 
\cite{cale}. This  approach views quantum dissipation as described by the 
linear coupling of the vortex coordinate to the coordinates of a bath of 
harmonic oscillators of prescribed dynamics. By means of this description 
of  quantum dissipation, we formulate our own approach to the dissipative 
tunneling of vortices subject to the fields ${\bf E}$ and ${\bf B}$ (the
magnetic-like field being treated in an effective way). 

The organisation of our article is as follows.  In Section II we sketch some 
basic facts about our Schwinger method for the relativistic vortex dynamics 
and show how the completely  non-relativistic limit yields a convenient 
formulation of the vortex tunneling problem. This is expanded in Section III,
where we treat in detail the situation in which dissipation and a Coulomb
interaction are present and dominate over vortex inertia. 
In Section IV we estimate the
tunneling rate for a range of reasonable values of the material parameters
and compare the results with previous estimates for the bulk nucleation 
rate. This Section contains also our conclusions. We work in the units 
system for which $\hbar=1$.

\vskip 1.0truecm

\section{ Vortex quantum dynamics:  a unified approach to vacuum decay 
through nucleation and tunneling }
\renewcommand{\theequation}{2.\arabic{equation}}
\setcounter{equation}{0}

The calculation that follows is the entirely non-relativistic version of our 
published QFT formulation \cite{ieju95} for the decay, via vortex-antivortex 
pair nucleation, of the ``vacuum'' represented by a supercurrent having 
number density ${\bf J}$ flowing in a superconducting thin film. Since the 
would-be particles should experience a force entirely analogous to the 
Lorentz force from a uniform electromagnetic field, we expect this vacuum 
to be unstable and decay via vortex pair creation.  We evaluate the 
probability amplitude for the vacuum decay in time ${\cal T}$,

\begin{equation} 
Z=\langle 0 | e^{ -i\hat{H}{\cal T} } | 0 \rangle {\equiv} e^{-i{\cal T}W_0}
\end{equation}

\noindent 
where $W_0={\cal E}(vac)-i\frac{\Gamma}{2}$ gives the energy of this
vacuum, ${\cal E}(vac)$, and its decay rate, $\Gamma$.  With a suitable 
normalization factor, the probability amplitude is given by a functional 
integral over field configurations. In the presence of a gauge field 
$A_{\mu}({\bf r},t)$ and external potential $V({\bf r})$, we have

\begin{eqnarray} 
Z&=&{\cal N}\int {\cal D}{\phi} \exp \left \{-i\int d^2r ~ dt ~ {\phi}^* 
\left ( -D_0^2+\frac{1}{\gamma}{\bf D}^2-{\cal E}_0^2 - V({\bf r}) 
\right ) {\phi} \right \}  \nonumber \\
&=&\exp \left \{ - Tr \ln \left ( -\frac{1}{\gamma}{\bf D}^2-D_3^2+{\cal E}_0^2 
+V({\bf r}) \right ) \right \}
\label{defin} 
\end{eqnarray}

\noindent 
which can be evaluated, in the Euclidean metric, by means of the identity

\begin{eqnarray} 
&&Tr \ln \frac{-D_E^2+{\cal E}_0^2+V({\bf r}) }{ \Lambda^2} 
\nonumber \\
&&= - \lim_{{\epsilon}{\rightarrow}0}\int_{\epsilon}^{\infty}
\frac{d\tau}{\tau} Tr \left \{ e^{-(-D_E^2+{\cal E}_0^2+V)\tau} -
e^{-{\Lambda}^2\tau} \right \} \label{identity} 
\end{eqnarray}

\noindent 
where (with $x_3{\equiv}it$) 
$D_E^2=D_3^2+\frac{1}{\gamma}(D_1^2+D_2^2)$ and where 
$D_{\mu}=\partial_{\mu} - i A_{\mu}$ is the usual covariant derivative. 
Notice that $\gamma=m/{\cal E}_0$, the ratio between the inertial mass 
$m$ and the activation energy ${\cal E}_0$, plays the role of the inverse 
square of the velocity of light. Thus the non-relativistic limit is implicitly
taken in the dissipation-dominated case, where $m{\rightarrow}0$.

The evaluation of the trace is straightforward by means of the Feynman 
path-integral.  We have, with a suitable normalization factor ${\cal N}(\tau)$

\begin{equation} 
Tr \left \{ e^{-(-D_E^2+{\cal E}_0^2+V)\tau} \right \} = {\cal N}(\tau) 
\int dq_0
\int_{q(0)=q(\tau)=q_0} {\cal D}q(s) e^{ -\int_0^{\tau} ds {\cal L}_E} 
\end{equation}

\noindent 
where the Euclidean version of the relativistic Lagrangian reads

\begin{equation} 
{\cal L}_E=\frac{1}{4} \dot{q}_t^2 + \frac{\gamma}{4} \dot{\bf q}^2 
- i \dot{q}_{\mu}A_{\mu}(q) + {\cal E}_0^2 + V({\bf q})
\label{relagrangian}
\end{equation}

\noindent 
Here $q$ is taken to be a $(d+1)$-dimensional coordinate describing the 
closed trajectories in the Euclidean space-time as a function of the 
Schwinger proper time $s$. We denote $q=\left(q_t(s),{\bf q}(s)\right)$
in terms of its time- and space-like components, respectively, the dot
representing the derivative $\dot{q}=\frac{dq}{ds}$. The part of the 
trajectory moving backward in time $q_t$ is therefore interpreted as 
describing the antivortex. The expression for the vacuum decay rate is 
therefore, quite generally

\begin{equation}
\frac{\Gamma}{2}=Im \int_0^{\infty} 
\frac{d\tau}{\tau} \int d{\bf q}_0 {\cal N}(\tau) \int_{q(0)=q(\tau)=q_0}  
{\cal D}q(s) e^{-\int_0^{\tau} ds {\cal L}_E}
\label{rerate}
\end{equation}

\noindent
where the integral over ${\bf q}_0$, the initial point in space of the closed 
paths, runs over the region of particle nucleation. In the case of 
vortex-antivortex pair creation in the bulk of the thin film, this region is the
surface of area $L^2$ of the film. 

In the problem considered in the present paper, vortices are already
present in a narrow strip of width $a{\approx}\xi$ (with $\xi$ the coherence 
length of the superconducting order parameter) along the edge of the 
sample where $\nabla \times {\bf J}$ is different from zero. Thus the
``vacuum'' in this case corresponds to a uniform distribution of vortices 
along the edge, whilst none of them is present in the bulk in the absence 
of an external magnetic field. The ``vacuum decay'' thus corresponds to the
possibility that some of the vortices are dragged away from the edge and
enter the bulk. We can still take Eq. (\ref{rerate}) as the starting point for 
determining the decay rate, provided we factor out the contributions coming 
from all paths describing particles propagating backwards in time $q_t$ 
and replace them with a suitable normalization factor. Also, since vortices 
are already present, a chemical potential $\mu^{*}$ will be introduced to 
cancel out their nucleation energy. We will take the non-relativistic limit 
explicitely, since now the advantage of the relativistic formulation (which in 
our case provided for a simultaneous description of both particles and 
antiparticles) is no longer useful.

We find that this way of approaching the tunneling problem, based on Eq.
(\ref{rerate}), has in our opinion a number of advantages over the standard
instanton-type calculation \cite{blatter,coleman}. The instanton (or WKB 
for point-like particles) calculation, although well established, nevertheless 
calls for the determination of the ``bounce'' solution of the path integral's 
saddle point equation and the functional integration of the fluctuations 
around it, a task often too hard to carry out explicitely. The present 
formulation based on an entirely non-relativistic Schwinger approach leads 
to a promising viable alternative, as we now show. The integration over  
the time paths $q_t(s)$ in Eq. (\ref{rerate}) is carried out by means of a 
sadde-point approximation which fixes the correct non-relativistic form of  
the Lagrangian. Our uniform-field situation corresponds to 
$A_{\mu}(q)=\frac{1}{2}F_{\mu\nu}q_{\nu}$, with the (2+1)-dimensional field 
tensor given by, after the analytic continuation to Euclidean time

\begin{equation} 
F_{\mu\nu}= \left ( \begin{array}{ccc} 0 & B & iE_x \\ 
-B & 0 & iE_y \\ -iE_x & -iE_y & 0 \end{array} \right )
\end{equation}

\noindent 
In the present treatment, the vortex chemical potential $\mu^{*}$ is also 
added to the scalar potential, $A_0\rightarrow A_0+\mu^{*}$, so that with 
a suitable choice the forward-moving vortices can be selected from the 
backward-moving antivortices. In our previous work \cite{ieju95} we have 
shown that the main role of  the magnetic-like field in the nucleation of 
vortex-antivortex pairs is to renormalise the friction coefficient (mimicked 
by the bath of harmonic oscillators included in the potential $V({\bf q})$),

\begin{equation}
\eta \rightarrow \eta_R=(\eta^2+B^2)/\eta
\end{equation}

\noindent
(denoted simply by $\eta$ in what follows), as well as the nucleation energy 

\begin{equation}
{\cal E}_0^2 \rightarrow  {\cal E}_{0R}^2={\cal E}_0^2 + \Delta {\cal E}(B)^2
\end{equation}

\noindent
where $\Delta {\cal E}(B)$ is a $B$- and $\eta$-dependent renormalization.
However, the friction coefficient $\eta$ is a rather uncertain normal-metal 
parameter and, moreover, we assume that there is an infinite reservoir of
vortices at the border of the film, their $\mu^{*}$ cancelling the value 
of the effective nucleation energy as well as all its renormalizations. 
Thus, we 
may safely neglect the effects of the magnetic-like component of the 
Magnus force and write the relativistic action in the form

\begin{eqnarray}
&&S_E(B=0)=\int_0^{\tau} ds \left \{ \frac{\gamma}{4}\dot{\bf q}^2
+\frac{1}{4}\dot{q}_t^2-\dot{q}_t({\bf E}\cdot{\bf q}+\mu^{*})+{\cal E}_{0R}^2
+V({\bf q}) \right \} \\
&&=\int_0^{\tau} ds \left \{ \frac{\gamma}{4} \dot{\bf q}^2 + \frac{1}{4}
\left ( \dot{q}_t-2({\bf E}\cdot{\bf q}+\mu^{*}) \right )^2 - \left (
{\bf E}\cdot{\bf q}+\mu^{*} \right )^2 + {\cal E}_{0R}^2 + V({\bf q})
\right \} \nonumber
\end{eqnarray}

\noindent
The saddle point approximation fixes the function $\dot{q_t}=dq_t/ds$ so 
as to have an extremum for the action: 
namely, we have the saddle-point condition

\begin{equation}
\frac{\delta S_E(0)}{\delta \dot{q}_t}=\frac{1}{2} \left ( \dot{q}_t
-2({\bf E}\cdot{\bf q}+\mu^{*}) \right ) = 0
\end{equation}

\noindent
from which we obtain  

\begin{equation}
\frac{dq_t}{ds}=2 ( {\bf E}\cdot{\bf q} + \mu^{*} ) \rightarrow 2\mu^{*}
\end{equation}

\noindent
where we take $\mu^{*}=\pm {\cal E}_{0R}$ to be the dominant energy 
scale. Here, the positive sign applies to particles moving forward in the 
time $q_t$ and the negative one to backward-moving antiparticles. The 
resulting saddle-point action for those trajectories moving forward up to the 
standard Euclidean time $T=2{\cal E}_{0R}\tau$ can therefore be taken as 
(indicating for short with $t$ the time component $q_t$)

\begin{equation}
S_E^{+}(0){\approx}\int_{0}^T dt \left \{ \frac{1}{2}m \left ( \frac{d{\bf q}}
{dt} \right )^2 - {\bf E}\cdot{\bf q} + U({\bf q}) \right \} {\equiv} S_{NR}
\end{equation}

\noindent
where the residual contribution in terms of the nucleation energy is 
cancelled by our choice for the chemical potential and where a term 
$({\bf E}\cdot{\bf q})^2/2{\cal E}_{0R}$ has been dropped. Also, we have 
denoted by $U({\bf q})=V({\bf q})/2{\cal E}_{0R}$ the non-relativistic  
potential energy. We now carry out the functional integral over the 
trajectories moving forward in time, replacing the contribution of the integral 
over the antivortex trajectories with some factor $\Phi$. This factor will also 
include the contribution arising from the fluctuations around the saddle 
point. The resulting expression for the tunneling rate through the barrier 
represented by the overall potential $U({\bf q})-{\bf E}\cdot{\bf q}$ becomes, 
from Eq. (\ref{rerate})

\begin{equation}
\frac{\Gamma}{2}{\approx}  Im 
\int_0^{\infty} \frac{dT}{T} \Phi {\cal N}_{NR}(T) \int_{{\bf q}(T)
={\bf q}(0)} {\cal D}{\bf q}(t) e^{-S_{NR}}
\label{tunnrate}
\end{equation}

\noindent
This expression contains only closed paths forward-moving in time $t$ 
and weighted by the non-relativistic action. The standard normalization 
factor ${\cal N}_{NR}(T)$ corresponds to the path integral for the 
non-relativistic free particle in $D=2$ dimensions, in the absence of
dissipation

\begin{equation}
{\cal N}_{NR}(T) \int_{{\bf q}(0)={\bf q}(T)} {\cal D}{\bf q}(t)
e^{ -\int_0^T dt \frac{1}{2} m \dot{\bf q}^2 } =\frac{m}{2\pi T}
\end{equation}

\noindent
To fix the overall factor $\Phi$, we proceed in the following manner. The 
expression for the non-relativistic vacuum decay amplitude is as follows

\begin{eqnarray}
\langle 0 | e^{-H_{NR} T} | 0 \rangle &=& {\cal N}_{NR}(T) \int {\cal D}
{\bf q} e^{ -S_{NR} } \nonumber \\
&=& \exp  - \left ( {\cal E}(vac)-i\frac{\Gamma}{2} \right ) T = 
e^{-{\cal E}(vac)T} \left ( \cos \frac{\Gamma T}{2} + i \sin 
\frac{\Gamma T}{2} \right )
\end{eqnarray}

\noindent
Since we are dealing with a highly-stable vacuum, we take the lowest 
order term in the expansion in powers of $\Gamma T$ for 
$\Gamma {\ll} {\cal E}(vac)$ to obtain

\begin{equation}
{\cal N}_{NR}(T) \int {\cal D}{\bf q} e^{-S_{NR}}=i\frac{\Gamma T}{2} 
e^{-{\cal E}(vac)T}+\cdots
\end{equation}

\noindent
which determines, inserted in Eq. (\ref{tunnrate}), the following 
expression for $\Phi$ 

\begin{equation}
\Phi={\cal E}(vac) =
\left \{ \int_0^{\infty} dT e^{-{\cal E}(vac)T } \right \}^{-1}
\end{equation}

\noindent
In deriving this expression we have taken into account the fact that the
calculation for the rate $\Gamma$ is based on a saddle-point approximation
where $\Phi$ is a function of the material's parameters only. The 
exponential $e^{-{\cal E}(vac)T}$ represents the vacuum probability 
amplitude in the absence of the electric field instability, thus we finally 
write

\begin{equation}
\frac{\Gamma}{2}=\frac{ Im \int_0^{\infty} \frac{dT}{T} {\cal N}_{NR}(T) 
\int {\cal D}{\bf q} e^{-S_{NR}} }{ \int_0^{\infty} dT {\cal N}_{NR}(T) 
\int {\cal D}{\bf q} e^{-S_{NR}^{(0)}} }
\label{nonrerate}
\end{equation}

\noindent
where the action $S_{NR}^{(0)}$ is the non-relativistic action in the 
absence of the electric field.  We have in this way reached a formulation
for the tunneling rate which leads to an expression in agreement with 
the standard WKB method. This can be checked by examples insofar as 
the exponential term $\beta$ is concerned, in the characteristic expression 
for the tunneling rate $\Gamma=\alpha e^{-\beta}$. 

\vskip 1.0truecm

\section{ Evaluation of the tunneling rate through the edge potential.
} \renewcommand{\theequation}{3.\arabic{equation}}
\setcounter{equation}{0}

In this Section we will apply the above general formalism to the situation
at hand, where the vortices tunnel from the edge strip under the combined
effect of the ``electric potential'' $-{\bf E}\cdot{\bf q}$ and the 2-D 
``Coulomb electrostatic potential''. The latter is due to the attraction
of the vortex by the edge, which is equivalent to the attraction by a 
virtual antivortex, like in the familiar virtual-charge method in standard
electrostatics. This Coulomb potential takes one of the equivalent forms 
$U_C(y)= K \ln (1+y/a)$ or 
$U_C(y)=\frac{1}{2} K \ln \left ( 1+(y/a)^2 \right )$, where $y$ is the 
distance from the edge, both acceptable extrapolations of the known
large-distance behavior. The coupling constant $K$ depends on the 
superfluid density $\rho_s^{(3)}$ and carrier's mass $m_0$ through the 
relationship \cite{aoth94} $K=2\pi\rho_s^{(3)}d/m_0$. However, its precise 
value in real materials is unknown and this parameter will be treated, like 
many others in this calculation, phenomenologically. 

When the Caldeira-Leggett quantum dissipation is taken into account
\cite{cale,ieju95}, we end up with an overall non-relativistic potential  
 
\begin{eqnarray}
U({\bf q})&=&\frac{1}{2{\cal E}_{0R}}V({\bf q})=U_C(y)+U_D({\bf q}) 
\nonumber \\ 
U_D({\bf q})&=&\sum_k \left \{ \frac{1}{2}m_k \dot{\bf x}_k^2 
+ \frac{1}{2}m_k\omega_k^2 \left ( {\bf x}_k + \frac{c_k}{2m_k\omega_k^2}
\right )^2 \right \} 
\end{eqnarray}

\noindent
with the oscillators' masses $m_k$ and frequencies $\omega_k$ 
constrained in such a way that the classical equation of motion for the 
resulting action reproduces the form (\ref{classic}). This requires the $c_k$ 
to satisfy the constraint (for the so-called ohmic case, which we consider)

\begin{equation}
\frac{\pi}{2}\sum_k\frac{c_k^2}{m_k\omega_k}{\delta}(\omega-\omega_k)
{\equiv}J({\omega})=\eta{\omega} 
\label{constr} 
\end{equation}

\noindent 
$\eta$ being the phenomenological friction coefficient of Eq. (\ref{classic})
(renormalised by $B$). Notice that this coefficient is unaffected by the
non-relativistic limit.  After a Fourier transformation in which the ${\bf x}_k$
modes can be integrated out \cite{ieju95}, we are lead to an effective 
non-relativistic action which in the dissipation-dominated 
($m \rightarrow 0$) regime reads, with ${\bf q}=(x,y)$

\begin{equation}
S_{NR}=2\pi \eta \sum_{n>0}n{\bf q}_n\cdot{\bf q}_n^{*} - ET \bar{y} + 
K \int_0^T dt \ln \left ( 1 + \frac{y(t)}{a} \right )
\label{fullact}
\end{equation}

\noindent
Here ${\bf q}(t)=\bar{\bf q}+\sum_{n{\neq}0} {\bf q}_n e^{i\omega_nt}$, 
where $\omega_n=2\pi n/T$, and furthermore $\bar{y}=\int_0^T dt y(t)/T$
stands for the $n=0$ mode in the $y$-direction. 
In order to 
render the problem tractable, at this point, we introduce a further
approximation for the interaction term of the action in Eq. (\ref{fullact}), by 
writing 

\begin{equation}
\int_0^T dt \ln \left ( 1+\frac{y}{a} \right ) {\approx} T \ln \left ( 1
+\frac{\bar{y}}{a} \right )
\label{approxnum}
\end{equation}

\noindent
which is justified in the limit of large exit times $T$ (see Appendix). We are 
therefore in a position to evaluate the numerator of the formula, Eq. 
(\ref{nonrerate}), for the tunneling rate, which we write schematically as
$\Gamma/2={\cal N}/{\cal D}$.

We define, for the numerator

\begin{equation}
{\cal N}{\equiv}Im \int_0^{\infty} \frac{dT}{T} {\cal N}_{NR}(T) 
\int {\cal D}{\bf q} e^{-S_{NR}}=Im \int_0^{\infty} \frac{dT}{T}
{\cal N}_{NR}(T) I_x I_y 
\end{equation}
 
\noindent
where the path integral over ${\bf q}(t)$ factorises into
$I_x=L\prod_{n=1}^{\infty} (1/2\eta n)$, representing the contribution of the 
free dissipative motion along the $x$-axis of the edge having length $L$, 
and in 

\begin{equation}
I_y=\int d\bar{y} I_G(\bar{y},T) e^{ET\bar{y} - KT \ln (1 + \bar{y}/a )}
\end{equation}

\noindent
the factor containing the dynamical effects, which we now evaluate. This we 
do most conveniently by integrating out first the real and imaginary parts, 
$Re ~ y_n=\psi_n$ and $Im ~ y_n=\xi_n$, of the modes $y_n$ for $n>0$. 
Since the path integral is restricted to loops which have their origin at 
$y(0)=y(T){\equiv}y_0=0$, the constraint 
$y_0=\bar{y}+\sum_{n{\neq}0}\psi_n=\bar{y}+2\sum_{n>0}\psi_n=0$ has 
to be imposed, corresponding to the vacuum wave-function 
$|\Psi_0(y_0)|^2{\sim}\delta (y_0)$.
This leads to the constrained 
Gaussian integrals 

\begin{eqnarray}
&&I_G(\bar{y},T)=\int \prod_{n=1}^{\infty} d\xi_n d\psi_n 
\delta \left ( \bar{y}+2\sum_{n=1}^{\infty}\psi_n \right ) e^{-2\pi\eta 
\sum_{n=1}^{\infty} n(\xi_n^2+\psi_n^2)}=\frac{1}{2} 
e^{-\frac{1}{2}\pi\eta \bar{y}^2} \times \nonumber \\
&& \times \prod_{n=1}^{\infty}  \left ( \frac{1}{2\eta n} \right )^{1/2} 
\int \prod_{n=2}^{\infty} d\psi_n \exp \left \{ - 2\pi\eta \sum_{n,m=2}^{\infty}
\psi_n(n\delta_{nm}+1)\psi_m - 2\pi\eta\bar{y}
\sum_{n=2}^{\infty}\psi_n \right \}
\end{eqnarray}

\noindent
where the $\delta$-function constraint has been taken care of through,
e.g., the $\psi_1$-integration. Introducing the matrix ($n,m=2,3,4\dots$)

\begin{equation}
M_{nm}=2\pi\eta (n\delta_{nm}+1)
\end{equation}

\noindent
the integrals can be evaluated, formally, to give

\begin{equation}
I_G(\bar{y},T)=\frac{1}{2\sqrt{\pi}}\prod_{n=1}^{\infty} \left ( 
\frac{\pi}{2\eta n} \right )^{1/2}  
\det M^{-1/2} e^{ -\frac{1}{2}\pi\eta \bar{y}^2 
+ \pi^2\eta^2\bar{y}^2\sum_{n,m=2}^{\infty} M_{nm}^{-1} }
\end{equation}

\noindent
Notice that the resulting expression involves infinite products and sums
that would lead to divergencies. However, since the dissipation is believed
to be suppressed above a characteristic material-dependent frequency
$\omega_c$, these products and sums are to be cut at a characteristic
integer $n^{*}=[\omega_c T/2\pi]$. Writing $M=2\pi\eta M_0(1+M_0^{-1}V)$, 
with $M_{0nm}=n\delta_{nm}$ and $V_{nm}=1$, we get

\begin{equation}
\det M=\prod_{n=2}^{n^{*}} (2\pi\eta n) \exp Tr \ln  (1+M_0^{-1}V)
\end{equation}

\noindent
The trace can be evaluated now by formally expanding $\ln (1+M_0^{-1}V)$,
to get

\begin{equation}
\det (1+M_0^{-1}V)=1+\sum_{n=2}^{n^{*}} \frac{1}{n}{\equiv}Q
\end{equation}

\noindent
In a similar way, we evaluate 
$M^{-1}_{nm}=\frac{1}{2\pi\eta} \left ( \frac{1}{n}\delta_{nm}-\frac{1}{Q nm} 
\right )$, yielding

\begin{equation}
\sum_{n,m=2}^{n^{*}}M_{nm}^{-1}=\frac{1}{2\pi\eta}(1-\frac{1}{Q})
\end{equation}

\noindent
from which we obtain

\begin{equation}
I_G(\bar{y},T)=\frac{1}{2} \left ( \frac{2\eta}{Q} \right )^{1/2} 
\prod_{n>0} \left ( \frac{1}{2\eta n} \right ) 
e^{-\frac{\pi\eta}{2Q}\bar{y}^2}
\end{equation}

\noindent
The numerator of our formula for the tunneling rate $\Gamma$ is therefore
the double integral

\begin{equation}
{\cal N}=Im \frac{1}{2} L \left ( \frac{2\eta}{Q} \right )^{1/2} \int_0^{\infty}
\frac{dT}{T} {\cal N}_{NR}(T) \prod_{n>0} \left ( \frac{1}{2\eta n} 
\right )^2 \int_0^{\infty} d\bar{y} e^{-\frac{\pi\eta}{2Q}\bar{y}^2+ET\bar{y}
-KT \ln (1+\bar{y}/a) }
\end{equation}

\noindent
We remark that in the above formula the following formal expression 
appears:
$\hat{N}{\equiv} {\cal N}_{NR}(T) \prod_{n>0} \left ( \frac{1}{2\eta n}
\right )^2$. We interpret this expression as an $m\rightarrow 0$ limit:

\begin{eqnarray}
\hat{N}&=&{\cal N}_{NR}(T) \prod_{n>0} 
\left ( \frac{\pi T^{-1}}{ m\omega_n^2+\eta\omega_n } \right )^2 
=\frac{m}{2\pi T} \prod_{n=1}^{n^{*}} \left ( 1+\frac{\eta T}{2\pi m}
\frac{1}{n} \right )^{-2} 
\nonumber \\ &&\rightarrow  \frac{\eta}{2\pi} \exp -T\Delta {\cal E}
\end{eqnarray}

\noindent
where $\Delta {\cal E}=\frac{\eta}{\pi m}(1+\ln (m\omega_c/\eta))$ is
a further renormalization of the activation energy which is also cancelled 
by tuning the vortex chemical potential $\mu^{*}$, so that, effectively,
$\hat{N}{\rightarrow}\frac{\eta}{2\pi}$.

\noindent
The integral is now evaluated by means of the steepest descent 
approximation in both $\bar{y}$ and $T$; the saddle-point equations are, 
with

\begin{equation}
S=\frac{\pi\eta}{2Q}\bar{y}^2-ET\bar{y}+KT \ln (1+\frac{\bar{y}}{a})
\end{equation}

\noindent
as the reduced action,

\begin{eqnarray}
\frac{\partial S}{\partial T}= - E\bar{y} + K \ln (1+\frac{\bar{y}}{a}) &=&0
\nonumber \\
\frac{\partial S}{\partial \bar{y}}=\frac{\pi\eta}{Q}\bar{y}-E\bar{T}+
K\bar{T}\frac{1}{a+\bar{y}} &=&0
\label{saddlept}
\end{eqnarray}

\noindent
the first of which yields $\bar{y}$ and the second $\bar{T}$, the mean exit
distance and time, respectively (since $n^{*}$ is a large cutoff integer,
$Q$ is kept fixed in the variation of the effective action). This leads to the 
expression $S_0=\pi\eta \bar{y}^2/2Q$ for the saddle-point effective action.
One has, furthermore, to integrate over the Gaussian fluctuations around 
$\bar{y}$ and $\bar{T}$. One can easily verify that there is a negative 
mode, thus giving an imaginary contribution, coming from the integration 
over the fluctuations in $T$. We in fact expand 

\begin{equation}
S=S_0+\alpha\delta y^2+\beta\delta y\delta T=S_0+\alpha (\delta y+
\frac{\beta}{2\alpha}\delta T)^2-\frac{\beta^2}{4\alpha}\delta T^2
\end{equation}

\noindent
with $\alpha=\frac{1}{2}(\frac{\pi\eta}{Q}-\frac{K\bar{T}}{(a+\bar{y})^2})$ and
$\beta=-E+\frac{K}{a+\bar{y}}$. This leads to the following evaluation of the 
numerator of Eq. (\ref{nonrerate})

\begin{eqnarray}
&&{\cal N}=Im \int_0^{\infty} \frac{dT}{T} {\cal N}_{NR}(T) 
\int {\cal D}{\bf q} e^{-S_{NR}} = \eta L \sqrt{ \frac{\eta}{2Q} }
\frac{ e^{-S_0} }{ \bar{T} \left ( E-K/(a+\bar{y}) \right ) }  \nonumber \\
&&S_0=\frac{\pi\eta}{2Q} \bar{y}^2
\label{num}
\end{eqnarray}

\noindent
where $Q=\ln [\omega_c\bar{T}/2\pi]$ for large enough $\bar{T}$.

We now come to the evaluation of the normalization denominator in
Eq. (\ref{nonrerate}). We remark that, the numerator being already of the
form $\alpha e^{-\beta}$, this denominator contributes only to the
specification of the prefactor $\alpha$ in the expression for the tunneling
rate $\Gamma$. To begin with, we carry out the Gaussian integrals over
$\xi_n$ and $\psi_n$. Here we use the alternative form of the Coulomb 
interaction, $U_C=\frac{1}{2}K\ln \left ( 1+(y(t)/a)^2 \right )$, and we 
expand the logarithm around the minimum (now at $y=0$) to lowest order 
(thus keeping all the integrals Gaussian). We define

\begin{equation}
{\cal D}{\equiv}\int_0^{\infty} dT {\cal N}_{NR}(T) \int {\cal D}{\bf q} 
e^{-S_{NR}^{(0)}}=\int_0^{\infty} dT {\cal N}_{NR}(T) I_x I_y^{(0)}
\end{equation}

\noindent
where $I_x$ is the same as for the numerator above, while 

\begin{eqnarray}
I_y^{(0)}&=&\int \prod_{n=1}^{\infty} d\xi_n d\psi_n \int d\bar{y}
\delta(\bar{y}+2\sum_{n=1}^{\infty}\psi_n) \times \nonumber \\
&\times & \exp \left \{ -2\pi\eta \sum_{n=1}^{\infty}
n(\xi_n^2+\psi_n^2)-\frac{KT}{2a^2}[2\sum_{n=1}^{\infty}(\xi_n^2+\psi_n^2)
+\bar{y}^2] \right \}
\end{eqnarray}

\noindent
Finally, repeating many of the steps of the calculation done for the 
Gaussian integrals of the numerator ${\cal N}$, we end up with

\begin{equation}
{\cal D}=\frac{a^2\eta^2}{K} L \int_0^{\infty} du \left ( 
1+2u Q(u) \right )^{-1/2} \prod_{n=1}^{n^{*}} \left ( 1+u/n \right )^{-1}
{\equiv} \frac{a^2\eta^2}{K} L ~~ \varphi ( \frac{\omega_c}{K} a^2\eta )
\label{den}
\end{equation}

\noindent
where $u=KT/2\pi a^2\eta$, $Q(u)=\sum_{n=1}^{n^{*}} 1/(n+u)$ and 
$\varphi$ is a dimensionless function of order unity. The evaluation of the 
tunneling rate is now formally completed, and an explicit analytic formula 
ensues from our (albeit approximate) treatment leading, from Eq.s 
(\ref{num}) and (\ref{den}), to the tunneling rate per vortex:

\begin{equation}
\frac{\Gamma}{2}={\cal N} / {\cal D}=\frac{K}{\varphi a^2 \sqrt{2Q\eta} }
\frac{e^{-S_0}}{\bar{T} \left ( E-K/(a+\bar{y}) \right )}
\end{equation}

\noindent
This expression is readily evaluated, once the parameters 
$a, \eta, \omega_c, K$ and $J$ are fixed, by evaluating numerically the
solutions of Eq. (\ref{saddlept}) and the integral in Eq. (\ref{den}) by
means of $n^{*}=[\omega_c T/2\pi]=[(\omega_ca^2\eta/K)u]$ ([] denoting 
the integer part). 
To obtain the tunneling rate per unit length, we must 
compute the density of vortices along the edge. Since 
$\rho_v=\frac{1}{2\pi} \nabla \times \nabla \theta$ is the bulk density in
terms of the condensate phase $\theta$, and since the number current is
${\bf J}=\frac{\rho^{(3)}_sd}{m_0}(\nabla \theta - {\bf A})$, we obtain

\begin{equation}
\int dy \rho_v = \frac{m_0}{2\pi \rho^{(3)}_sd} \int dy 
\partial_y J_x = J/K
\end{equation}

\noindent
The final expression for $R$, the tunneling rate per unit length, is thus

\begin{equation}
R=\frac{J}{K} \Gamma 
\end{equation} 

We conclude this Section by noticing that, since $E=2\pi J$ in terms of the 
supercurrent density, and since the saddle-point equation (\ref{saddlept}) 
has iterative solution 
$\bar{y}=\frac{K}{E}\ln (1+\bar{y}/a)=\frac{K}{E} \{ \ln (1+K/Ea)+\cdots \}$,
we can cast the dominant exponential factor determining $\Gamma$ in
the form

\begin{equation}
S_0=\frac{ \eta \{ K\ln (1+K/2\pi Ja)+\cdots \}^2 }{ 8\pi J^2 Q }
\end{equation}

\noindent
This yields almost the same leading dependence (logarithmic corrections
apart) $\Gamma {\approx} e^{-(J_0/J)^2}$ on the external current that was 
also obtained for the homogeneous bulk-nucleation phenomenon 
\cite{ieju95} (in terms of the tunneling length $\ell_T{\sim}K/E$ we could
also rewrite $\Gamma{\sim}e^{-\eta\ell_T^2}$). Indeed, the effective 
Coulomb energy $K\ln (1+K/2\pi Ja)$ can be interpreted as playing the role 
of the activation energy ${\cal E}_0$ renormalised by the vortex-antivortex 
interactions.

\section{ Discussion and Estimate of the Tunneling Rate.
} \renewcommand{\theequation}{4.\arabic{equation}}
\setcounter{equation}{0}

We have presented a new treatment for the quantum tunneling of  vortices 
from the boundary of a thin superconducting film. The formulation makes 
use of the idea that tunneling can be viewed as a pair-creation process in
which only forward-moving time trajectories contribute to the Schwinger
path integral. Backward-moving antivortices represent contributions to the
path integral that are factorised out, and we have made use of the fact that 
an infinite reservoir of vortices is present at the edge. This leads to a 
mathematically convenient formulation of the evaluation of the tunneling
rate, which we have worked out in detail by assuming that the vortices 
experience an attractive 2D Coulomb-like potential confining them to the
edge. By means of some approximate treatment of the logarithmic
Coulomb interaction, and of the saddle-point evaluation of the path integral,
we have reduced all integrations to Gaussian integrals which afford a closed
analytical expression for the tunneling rate. This can then be estimated
by chosing material parameters suitable for typical cuprate superconducting
systems, e.g. YBCO films, that afford some of the highest critical current 
densities ($J_c{\sim}10^7$ Acm$^{-2}$ at 77 K \cite{cypa}). 
To fix the indicative value of the friction coefficient $\eta$, we make use 
of the the Bardeen-Stephen formula \cite{bast} linking the upper critical 
field $B_{c2}$ to the normal metal resistivity $\rho_n$:
$\eta=\Phi_0 B_{c2}/\rho_n$ ($\Phi_0$ being the flux quantum). This 
yields $\eta{\approx}10^{-2}$ $\AA^{-2}$ for YBCO. A rather uncertain
parameter is the cutoff frequency $\omega_c$, although the dependence
on it of the tunneling rate is rather weak. We therefore take indicatively
$\omega_c{\approx}80$ K, just under the expected value of the
Kosterlitz-Thouless transition temperature in these films \cite{reviews}.
As for the edge thickness, we take $a{\approx}\xi{\approx}10 \AA$ of
the order of magnitude of the coherence length. Finally, there is the
value of the Coulomb interaction strength, $K$. Since we must have
$K<\omega_c$, we present in Fig. 1 the dependence of the tunneling
rate on the current density for some indicative values of $K$. The effect of 
temperature is taken into account phenomenologically \cite{ieju95}, by
introducing thermal current fluctuations in the exponential of the formula
for the rate, $J^2\rightarrow J^2+\Delta J(T)^2$. The fluctuation term 
$\Delta J(T)$ is fixed by requiring that for $J=0$ an Arrhenius form 
$\Gamma\propto e^{-U_{max}/T}$ is obtained for the rate, with  the 
barrier's height $U_{max}=K \ln (K/2\pi Ja) - K+2\pi Ja$ as the activation 
energy.  Notice that the barrier is infinite for $J{\rightarrow}0$, so that 
there is no zero-current tunneling of vortices even for non-zero temperature.
It can be seen from Fig. 1 that the effect of thermal fluctuations, as 
estimated from the above interpolative assumption, is indicatively to raise 
the tunneling rate by some orders of magnitude.

The results are summarised in Fig. 1, where it is shown that tunneling of 
vortices from the film's boundary can become a highly likely phenomenon 
for sufficiently large current densities $J$ and relatively low Coulomb 
couplings $K$. The process is aided by the presence of thermal current 
fluctuations. For even larger currents than those considered here, the 
process could be described by means of a classical treatment based on the
time-dependent Ginzburg-Landau equation \cite{argish}. 
It is interesting to compare the results obtained in Fig. 1 with 
the estimates carried out for the bulk nucleation process \cite{ieju95}. 
There, the maximum nucleation rates observed (also taking temperature into 
account) were in the region of $\approx$ $10^{13}$ $\mu$m$^{-2}$s$^{-1}$ 
for a current $J=10^7$ A cm$^{-2}$ and assuming that pinning centers act as 
nucleation seeds. We therefore conclude that edge tunneling appears to 
be a more likely supercurrent decay mechanism for normal 
superconducting sample geometries. Yet, the bulk-nucleation of 
vortex-antivortex pairs is also a competitive parallel mechanism that could 
be singled-out by means of an appropriate choice of sample geometry.

Although the measurements of the residual resistence in a superconductor 
may not allow to distinguish between these two competing mechanisms, 
we stress that from the point of view of vortex-tunneling microscopy the two 
processes remain entirely separate observable phenomena. Bulk 
pair-nucleation remains, in particular, a completely open challenge for 
observation of a phenomenon that even in the context of standard QED has 
remained, to our knowledge, so far elusive for static fields.

\begin{center} {\bf Acknowledgements } \end{center}

One of us (G.J.)  is grateful to the International School for Advanced 
Studies in Trieste, where part of his work was carried out, for hospitality 
and use of its facilities.

\section*{ Appendix }
\renewcommand{\theequation}{A.\arabic{equation}}
\setcounter{equation}{0}

In this Appendix, we justify the use of the approximation, Eq. 
(\ref{approxnum}), used in our calculation for handling the ``Coulomb'' 
interaction in the path integral of the numerator of our formula for the
nucleation rate. For large $\bar{y}$, and with $\omega_n=2\pi n/T$:

\begin{equation}
\int_0^T dt \ln \left ( 1 + \frac{y(t)}{a} \right ) \approx T \ln \left ( 1 + 
\frac{\bar{y}}{a} \right ) + \sum_{l=1}^{\infty} \frac{(-)^{l+1}}{l} \int_0^T dt
\left ( \sum_{n{\neq}0} \frac{y_n}{\bar{y}} e^{i\omega_nt} \right )^l
\end{equation}

\noindent
At the saddle point, $y_n=-\bar{y}/2nQ$ (as can be easily verified by
varying Eq. (\ref{fullact})) and therefore we have, for example

\begin{equation}
\int_0^T dt \left ( \sum_{n{\neq}0} \frac{y_n}{\bar{y}} e^{i\omega_nt}
\right )^2=\frac{T}{4Q^2}\sum_{n=1}^{\infty} \frac{1}{n^2}
\end{equation}

\noindent
Since $Q=\ln [\frac{\omega_cT}{2\pi}]$ and $T{\sim}\frac{\eta K}{E^2}$,
we get the estimate

\begin{equation}
\int_0^T dt \ln \left ( 1 + \frac{y(t)}{a} \right ) \approx T \left [ 
\ln \left ( 1 + \frac{\bar{y}}{a} \right ) + O \left ( \ln \frac{\eta\omega_cK}
{E^2} \right )^{-2} \right ]
\end{equation}

\noindent
Thus the approximation of taking the ``Coulomb'' potential of the form 
$K \ln ( \bar{y}/a )$ is justified for very low currents 
($E=2\pi J{\rightarrow}0$). We still keep this form, when the current is
not so low, as we believe it captures the essential point of the dynamical
picture.

\begin{center} FIGURE CAPTIONS \end{center}

\noindent
Figure 1. Plot of the tunneling rate per unit length, $R$, as function of
the supercurrent 
density, $J$, for the values $K=30$ K and $K=50$ K. Dashed lines,
$T=0$ K; full lines, $T=2.5$ K.
  
\vfill

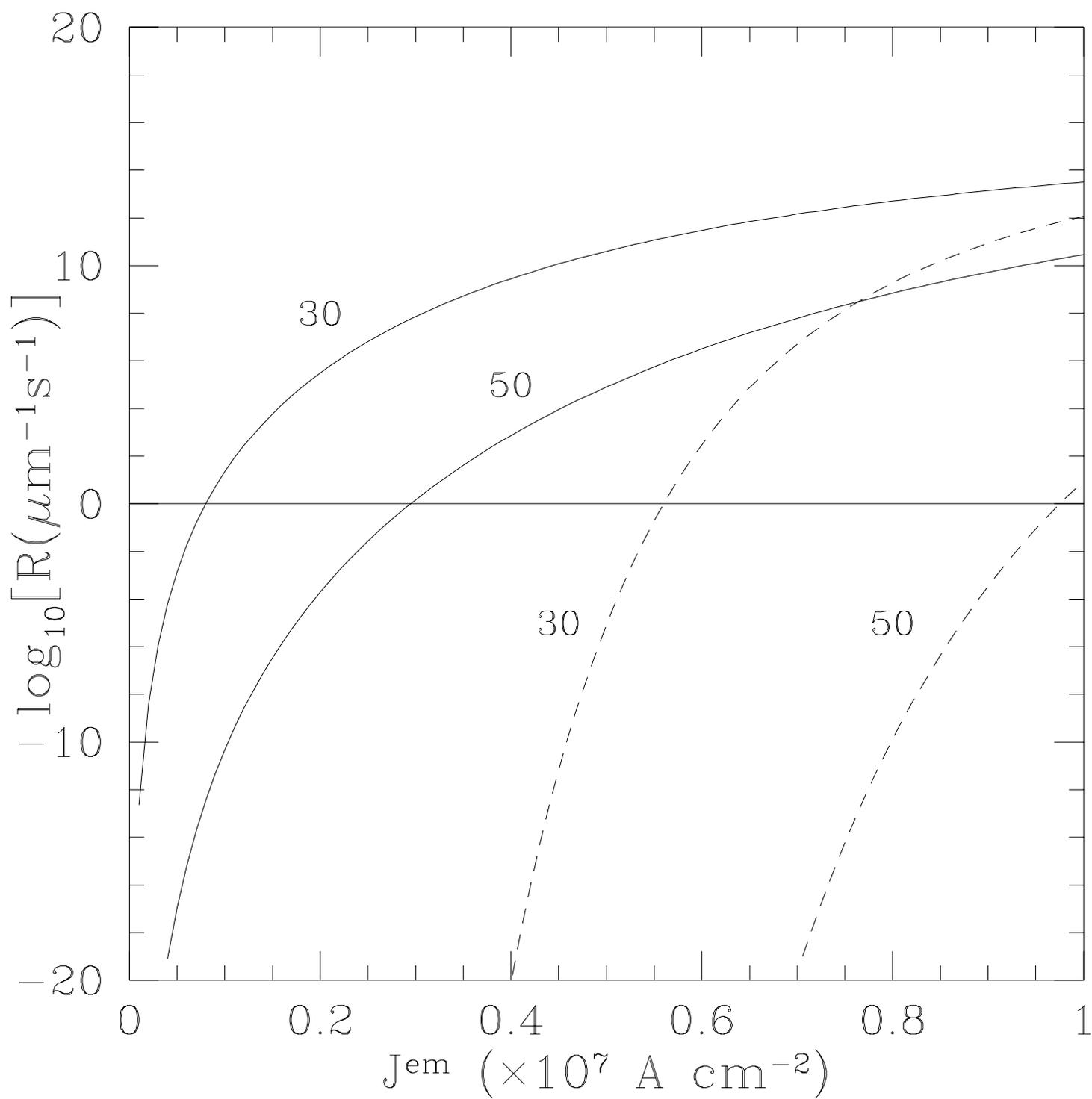
\end{document}